\documentclass[journal,twoside,letterpaper]{IEEEtran}
\usepackage{graphicx}
\usepackage{dblfloatfix}    % To enable figures at the bottom of page
\usepackage{xcolor}
\usepackage[compress]{cite}
\usepackage{psfrag}
\usepackage{subfig}
\usepackage{multirow}
\usepackage[cmex10]{amsmath}
\usepackage{amsfonts}
\usepackage{amssymb}

\begin{document}

\title{A Clock Synchronizer for Repeaterless Low Swing On-Chip Links}
\author{Naveen Kadayinti, 
Maryam Shojaei Baghini ~\IEEEmembership{Senior Member,~IEEE} and 
Dinesh K. Sharma ~\IEEEmembership{Senior Member,~IEEE}%
\thanks{The authors would like to thank  Mahima Arrawatia (IIT Bombay) and 
Amrith Sukumaram (IIT Madras) for their help during the tapeout. 
The authors are also thankful to Tata Consultancy Services (TCS)
and the SMDP programme of the Government of India for student scholarships
and for providing funds for EDA tools respectively.}
\thanks{The authors are with the Department of Electrical 
Engineering at the Indian Institute of Technology Bombay, 
Mumbai 400076, India. (email : 
\mbox{naveen@ee.iitb.ac.in}, 
\mbox{mshojaei@ee.iitb.ac.in}, 
\mbox{dinesh@ee.iitb.ac.in}).}
}

%\markboth{IEEE TRANSACTIONS ON CIRCUITS AND SYSTEMS-I: REGULAR PAPERS}%
%{Kadayinti \MakeLowercase{\textit{et al.}}: A Clock Synchronizer for Repeaterless Low Swing On-Chip Links}

\maketitle
\begin{abstract}
A clock synchronizing circuit for repeaterless low swing interconnects 
is presented in this paper. The circuit uses a delay locked loop (DLL) 
to generate multiple phases of the clock, of which the one closest to 
the center of the eye is picked by a phase detector loop. The picked 
phase is then further fine tuned by an analog voltage controlled delay 
to position the sampling clock at the center of the eye. A clock domain 
transfer circuit then transfers the sampled data to the receiver clock 
domain with a maximum latency of three clock cycles. 
The proposed synchronizer has been designed and fabricated in 130 nm 
UMC MM CMOS technology. The circuit consumes \mbox{1.4 mW} 
from a 1.2 V supply at a  data rate of 
\mbox{1.3 Gbps}. Further, the proposed synchronizer 
has been designed and simulated in TSMC 65 nm CMOS technology. 
Post layout simulations show that the synchronizer consumes \mbox{
1.5 mW} from a 1 V supply, at a data rate of \mbox{4 Gbps} in this 
technology.
\end{abstract}

\begin{IEEEkeywords}
Current mode interconnects, Low swing interconnect, repeater insertion, 
clock data recovery, Mesochronous synchronizers.
\end{IEEEkeywords}

\section{Introduction}
\label{sec:intro}
\IEEEPARstart{I}{t} has been well established that the performance of
digital processing systems is limited by the throughput and power
consumption of global interconnects
\cite{Katoch-esscirc-2005,Rho-isscc-2007,Mensink-isscc-2007,Kim-isscc-2009}.
Repeater insertion alleviates this
problem to some extent, but it increases the power consumption of the link,
while also bringing in additional constraints in placement and routing.
These limitations
have resulted in a lot of interest in repeaterless low swing interconnects, with
equalization at the transmitter \cite{kim-jssc2010, Rho-isscc-2010, naveen_vlsi13}  
or receiver \cite{Tx_rx_codesign} or both \cite{Lee-jssc14}, to improve the speed,
while keeping the power consumption low. The receiver circuit for low swing
interconnects is a comparator which converts the received low swing signal
to CMOS levels. To keep the power consumption and latency low, regenerative
clocked comparators are used \cite{Mensink-jssc-2010,naveen_vlsi13,Lee-issc13}.

While low swing interconnects help maximize the throughput of long
interconnects, the latency of the interconnects is still high and can
be as high as multiple cycles. This is the case even for transmission line based
interconnects that operate at the theoretical minimum latency \cite{richard-jssc}.
Conventional repeater inserted interconnects use
synchronizing flip-flops at section lengths less then the critical path
delay of the full chip to maintain synchronization \cite{Lu-DATE-2002}. 
Inserting synchronizing flip-flops along a low swing line
will however mean that the improvements offered by repeaterless links
will not be leveraged to full potential.
Hence, a clock re-timing circuit that ensures that the data is sampled at the 
center of the eye is required. Further, the resolved data should be 
transferred to 
the receiver clock domain with an appropriate synchronizer.
This problem, despite being serious, has not received much attention in the literature.

\begin{figure}[!h]
\psfrag{CkTx}{\small{$\phi_{Tx}$}}
\psfrag{CkRx}{\small{$\phi_{Rx}$}}
\psfrag{n+alpha}{\small{$(n+\alpha)\times T$}}
\psfrag{Ckd}{\small{$\phi_d$}}
\psfrag{Source}{\small{Source}}
\psfrag{Sense}{\small{Sense}}
\psfrag{Amplifier}{\small{Amplifier}}
\psfrag{Clock}{\small{Clock}}
\psfrag{Domain}{\small{Domain}}
\psfrag{Transfer}{\small{Transfer}}
\psfrag{Destination}{\small{Destination}}
\psfrag{Retiming}{\small{Clock Retimer}}
\psfrag{Line}{\small{Interconnect}}
\centering
\includegraphics[width=8cm]{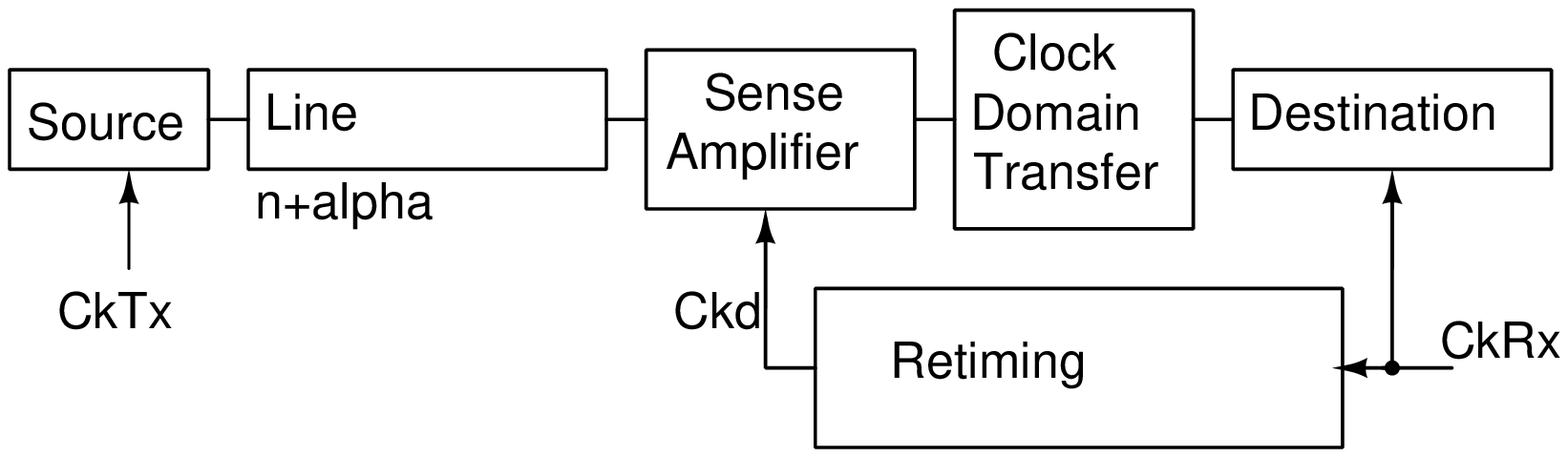}
\caption{Block diagram of a repeaterless low
swing interconnect system. \mbox{$\phi_{Tx}$: Transmitter clock}, 
\mbox{$\phi_{Rx}$}: 
Receiver clock, \mbox{$\phi_d$}: retimed sampling clock, 
($(n+\alpha)\times T$): Repeaterless interconnect delay, where
\mbox{$n \in \mathbb{Z}^{\geq 0}$},
\mbox{$\alpha \in \mathbb{R}$} \& \mbox{$\alpha < 1$}, 
\mbox{$T$: system clock period.}}
\label{fig:CPA}
\end{figure}

Fig. \ref{fig:CPA} shows a block diagram of a typical low swing interconnect system.
Here, $\phi_{Tx}$ is the transmitter clock, $\phi_{Rx}$ is 
the receiver clock,
and $\phi_d$ is the desired sampling clock with its active edge at the center of
the data eye. The delay of the repeaterless interconnect, that must be 
compensated for, is expressed as
$(n+\alpha)\times T$, where $T$ is the system clock period, $n$ is a positive
 integer (including 0),
and $\alpha$ is a positive real number which is less than 1. 
Mensink et al. in \cite{Mensink-jssc-2010} estimate
the delay using extracted simulations of the interconnect and add
appropriate delay at design time itself. This needs sufficient over design
so as to ensure proper operation at the desired frequency across process
corners. Source synchronous schemes, akin
to the one described in \cite{vivekde-tcas-09}, are not compatible with low swing interconnects
as converting the low swing clock to full swing clock will need buffers, whose
delay again cannot be predicted accurately at design time. 

\begin{figure*}[t!]
\psfrag{CkTx}{\small{$\phi_{Tx}$}}
\psfrag{CkRx}{\small{$\phi_{Rx}$}}
\psfrag{n+alpha}{\small{$(n+\alpha)\times T$}}
\psfrag{Ckd}{\small{$\phi_d$}}
\psfrag{Source}{\small{Source}}
\psfrag{Interconnect}{\small{Interconnect}}
\psfrag{Clock}{\small{Clock}}
\psfrag{Domain}{\small{Domain}}
\psfrag{Transfer}{\small{Transfer}}
\psfrag{Destination}{\small{Destination}}
\psfrag{Negative}{\small{Negative Feedback}}
\psfrag{Ckin}{\small{$Ck_{in}$}}
\centering
\includegraphics[width=15cm]{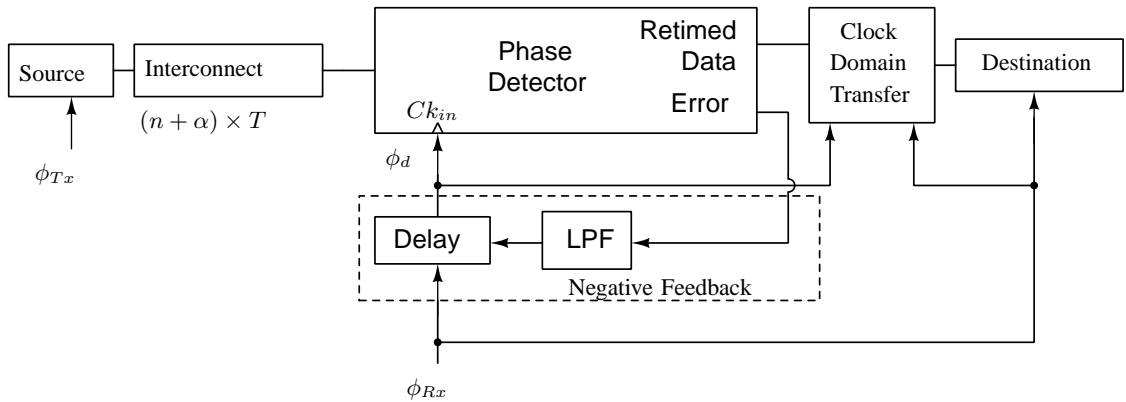}
\caption{Conceptual block diagram of the clock
recovery system at the receiver of a low swing interconnect system. 
\mbox{LPF - Low pass filter.}}
\label{fig:CPA_concept}
\end{figure*}

Clock and data recovery (CDR) circuits have been reported 
for off chip interconnects, and 
are well known \cite{CDR_razavi}. These circuits are typically phase locked loops
that lock a local oscillator's frequency to the incoming data frequency.
However, for on-chip interconnects, only the phase needs to be recovered
and a clock running at the correct frequency is available at the receiver.
Fig. \ref{fig:CPA_concept} shows the concept of such a clock recovery circuit.
Here a phase detector senses the phase error between the data and the clock
and generates an error signal. The integrated error signal controls a
delay circuit that delays the sampling clock. The negative feedback 
minimizes the error, resulting in the sampling clock position at the
center of the eye.
Lee et al.
in \cite{Lee-jssc14} report a source synchronous link that uses 
a digitally controlled delay line in the clock path at the receiver, which is 
trained in calibration
mode before the interconnect is used. 
For calibration, 0.5 unit interval delay is inserted at the transmitter 
and a pattern toggling between 1 and 0 is sent. 
The receiver clock delay is then swept
to find the crossing over of the data and clock edges,
and after completion of training, the 0.5 unit interval delay is removed.
This however means that
the technique cannot be extended to adaptive synchronizers.
Also, the accuracy is limited by the phase quantization error.
If a conventional phase detector like the Alexander phase detector is used,
the delay line should be initialized to the center of the range,
due to its finite range.
While infinite phase delay can be accomplished using phase interpolators
as reported in \cite{phase_interpolator}, the circuit is
predominantly analog and complex,
making it likely to suffer from mismatch in scaled technologies.
Another limitation of analog circuits is that their state cannot 
be easily saved for fast initialization on subsequent power ups. 

In this work, we present a fast automatic synchronizer
that is adaptive and does not have the limitations mentioned above.
The circuit is built around a DLL
that generates multiple phases of the clock.
A phase detector loop picks the DLL phase closest to the center of the eye.
In order to reduce the phase quantization error,
an analog delay line is used to fine tune the selected clock phase
so as to position it at the center of the data eye. 
A similar
concept has been used for clock synthesis in \cite{4vcdl,dualDLL},
in which a coarse DLL generates multiple phases of the clock, which
are delayed by multiple delay lines or interpolated to obtain the
clock of the correct phase. However these techniques are reported
for clock synthesis and not for clock and data recovery, and either use 
multiple VCDL's \cite{4vcdl} or use digital fine tuning which 
causes jitter due to dithering. The proposed technique uses a much simpler
implementation with only one VCDL.

Once sampled, the data must then be transferred to the receiver clock domain.
Serial synchronizers with two or more flip-flops are typically used
for such clock domain transfers \cite{Dally},
which however bring a penalty in latency.
In the design described here, using one of the phases of the DLL,
a low latency clock domain transfer to the receiver clock domain
has been implemented.

The paper is organized as follows. Section \ref{sec:cdr_arch} 
describes the architecture of the clock synchronizer.
Clock domain transfer, from the sampling clock domain to the
receiver clock domain, is described in section \ref{sec:cdt}. Jitter
analysis is presented in section \ref{sec:jitter} which is
followed by a discussion on implementation and results
in Section \ref{sec:implementation}.
Section \ref{sec:conclude} concludes this paper.
Appendix \ref{app:APD} discusses an anomaly
in DLL based clock recovery circuits using the Alexander phase detector.

\section{Clock recovery circuit}
\label{sec:cdr_arch}
\begin{figure*}
\centering
\psfrag{CkRx}{\small{$\phi_{Rx}$}}
\psfrag{Ckd}{\small{$\phi_d$}}
\psfrag{Data}{\small{Data}}
\psfrag{Input}{\small{Input}}
\psfrag{Phase}{\small{Phase}}
\psfrag{Detector}{\small{Detector}}
\psfrag{Charge}{\small{Charge}}
\psfrag{Pump}{\small{Pump}}
\psfrag{(weak)}{\small{(weak)}}
\psfrag{(strong)}{\small{(strong)}}
\psfrag{Logic}{\small{Logic}}
\psfrag{UPDN}{\small{UP DOWN}}
\psfrag{UP}{\small{$UP$}}
\psfrag{UPst}{\small{$UP_{strong}$}}
\psfrag{DNst}{\small{$DN_{strong}$}}
\psfrag{DN}{\small{$DN$}}
\psfrag{Counter}{\small{Counter}}
\psfrag{Divider}{\small{Divider}}
\psfrag{Switch Matrix}{\small{Switch Matrix}}
\psfrag{DLL}{\small{DLL}}
\psfrag{VCDL}{\small{VCDL}}
\psfrag{Window}{\small{Window}}
\psfrag{Comparator}{\small{Comparator}}
\psfrag{Retimed Data}{\small{Retimed Data}}
\psfrag{UD}{\small{$\overline{UP}/DN$}}
\psfrag{EN}{\small{$Enable$}}
\psfrag{VC}{\small{$V_{c}$}}
\psfrag{Qs}{\small{$Q_0-Q_9$}}
\psfrag{Fine}{\small{Fine tuning loop}}
\psfrag{Coarse}{\small{Coarse tuning loop}}
\psfrag{phin}{\small{$ $}}
\subfloat[The Synchronizer system]{\includegraphics[width=14cm]{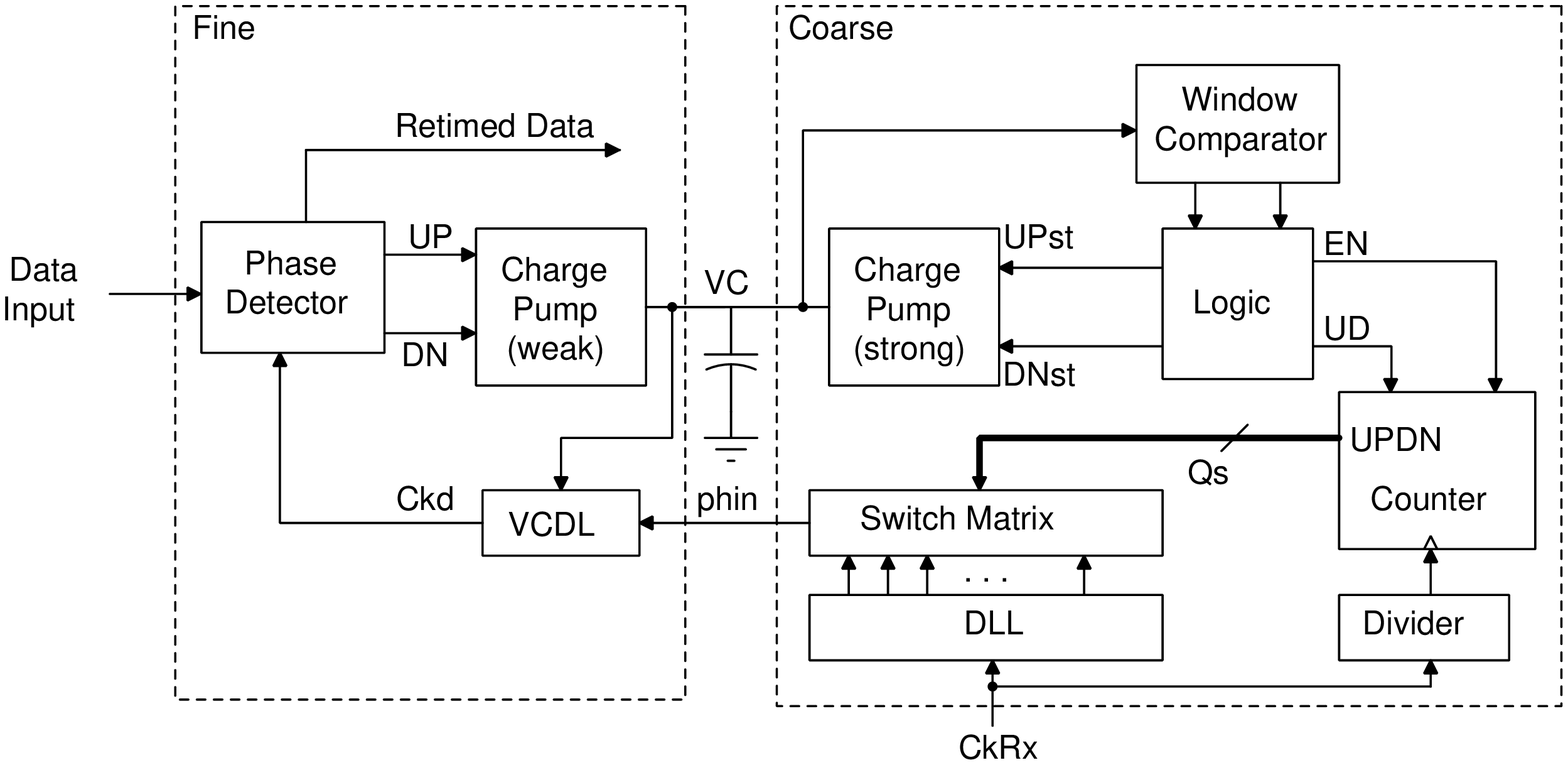}
\label{fig:CPA_analog}}
\\
\centering
\psfrag{Q}{\small{$Q$}}
\psfrag{D}{\small{$D$}}
\psfrag{CK}{\small{$CLK$}}
\psfrag{Q0}{\small{$Q_0$}}
\psfrag{Q1}{\small{$Q_1$}}
\psfrag{Q2}{\small{$Q_2$}}
\psfrag{Q8}{\small{$Q_8$}}
\psfrag{Q9}{\small{$Q_9$}}
\psfrag{Qi-1}{\small{$Q_{i-1}$}}
\psfrag{Qi+1}{\small{$Q_{i+1}$}}
\psfrag{Qi}{\small{$Q_i$}}
\psfrag{Preset}{\small{Preset}}
\psfrag{Reset}{\small{Reset}}
\psfrag{CLR}{\tiny{CLR}}
\psfrag{MUX}{\tiny{MUX}}
\psfrag{ENABLE}{\small{$Enable$}}
\psfrag{UPDN}{\small{$\overline{UP}/DN$}}
\subfloat[UP DOWN counter]{\includegraphics[width=14.5cm]{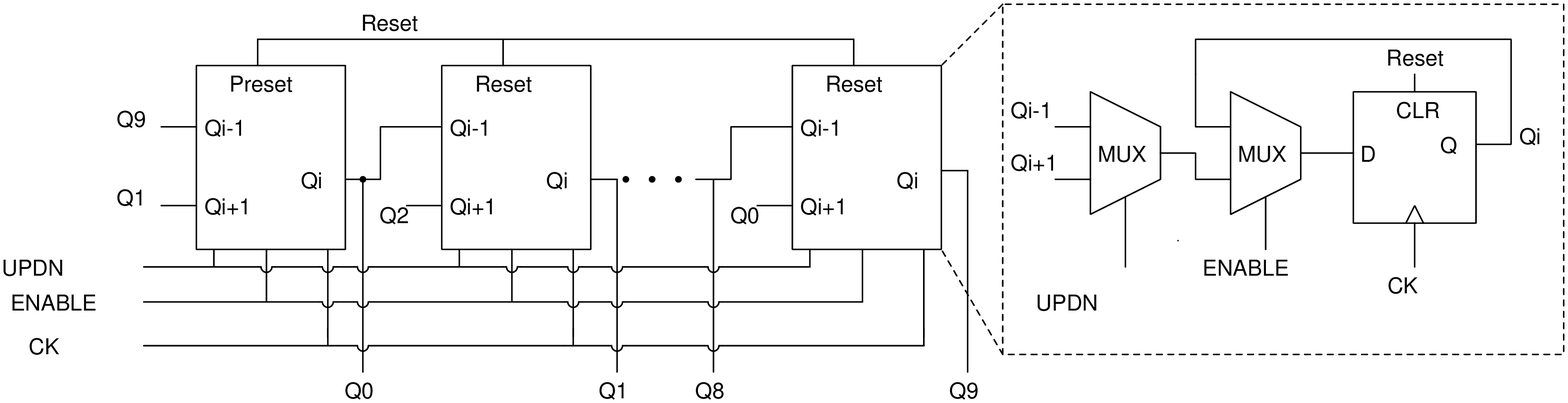}
\label{fig:ring_counter}}
\caption{(a) Block diagram clock synchronizer system, divided
into fine tuning and coarse tuning loops.
\mbox{VCDL - voltage controlled delay line},
\mbox{$Vc$ - control voltage}, and (b) Schematic of the UP DOWN ring counter used in the synchronizer.}
\label{fig:CPA_system}
\end{figure*}

Fig. \ref{fig:CPA_system}\subref{fig:CPA_analog} shows a block diagram
of the proposed clock synchronizing circuit.
It consists of a coarse tuning loop and a fine tuning loop.
The main component of fine tuning loop is a voltage 
controlled delay line (VCDL)
that provides a controllable delay to the clock.
The main component of the coarse tuning loop is a DLL
that generates multiple phases of the clock,
of which the one closest to the center of the data eye
is picked by the control loop.
Since the system is of the first order,
the loop filter is a single capacitor,
as shown between the fine and coarse tuning loops in 
Fig. \ref{fig:CPA_system}\subref{fig:CPA_analog}. 

Operation of the circuit starts
with the fine tuning loop trying to get the clock
to the center of the data eye, by delaying the clock using the
VCDL.
If the entire VCDL range is spent before lock is achieved (which
is identified by the control voltage exceeding preset bounds),
the coarse tuning loop is woken up
to pick the next phase of the DLL as the source clock
and the control voltage is reset to lie within the window.
This process repeats until the phase closest to the center
of the eye is selected and the VCDL range is sufficient
to lock the clock to the exact center of the eye.
Referring to Fig. \ref{fig:CPA_system}\subref{fig:CPA_analog},
the input low swing data first goes to a phase detector.
The $UP$ and $DN$ pulses from the phase detector
are averaged using a weak charge pump.
The averaged control voltage ($V_c$)
then modulates the delay of the VCDL.
The VCDL source clock comes from the coarse tuning loop, 
and is one of the phases of the DLL.
The VCDL is designed to have a range greater than 1 phase step of the DLL.
 
In the coarse tuning loop,
a window comparator senses the control voltage ($V_c$)
and if it exceeds a predetermined range
(which is the range of the control voltage for the VCDL
in the fine tuning loop), it triggers the logic block.
The logic block is an FSM that
generates the $Enable$ and $\overline{UP}/DN$ signals
for the one hot ring counter (shown 
in Fig. \ref{fig:CPA_system}\subref{fig:ring_counter}).
The state of the ring counter is used
to select one of the phases of the clocks generated by the DLL.
The direction in which the ring counter counts
depends on whether the control voltage exceeds the upper threshold
or is below the lower threshold of the window comparator.
Whenever the window thresholds are crossed,
a secondary strong charge pump resets the control voltage ($V_c$)
to bring it within the window.
When the control voltage is within the window thresholds,
the digital circuits retain their state by de-asserting the
$Enable$ signal.

\begin{figure*}[h]
\centering
\psfrag{Vc__}{\small{$V_c$}}
\psfrag{VH__}{\small{\hspace{-1ex}$V_H$}}
\psfrag{VL__}{\small{$V_L$}}
\psfrag{Timemicros}{\small{Time ($\mu s$)}}
\psfrag{Voltage}{\small{Voltage (V)}}
\subfloat[Control voltage]{\includegraphics[width=7cm]{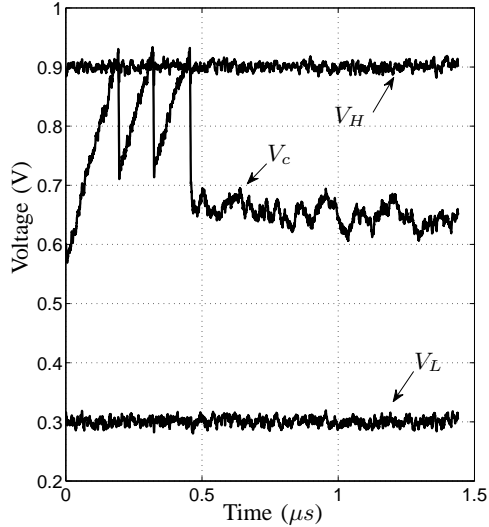}\label{fig:Vc_alex}
\label{fig:vc}}
\hspace{5ex}
\subfloat[Ring counter state]{
\psfrag{Q0_}{\small{$Q_0$}}
\psfrag{Q1_}{\small{$Q_1$}}
\psfrag{Q2_}{\small{$Q_2$}}
\psfrag{Q3_}{\small{$Q_3$}}
\psfrag{Q4_}{\small{$Q_4$}}
\psfrag{Q5_}{\small{$Q_5$}}
\psfrag{Q6_}{\small{$Q_6$}}
\psfrag{Q7_}{\small{$Q_7$}}
\psfrag{Q8_}{\small{$Q_8$}}
\psfrag{Q9_}{\small{$Q_9$}}
\psfrag{UPDN____}{\small{$\overline{UP}/DN$}}
\psfrag{Enable__}{\small{$Enable$}}
\psfrag{UPStrong__}{\small{$UP_{strong}$}}
\psfrag{DNstrong__}{\small{$DN_{strong}$}}
\includegraphics[width=7cm]{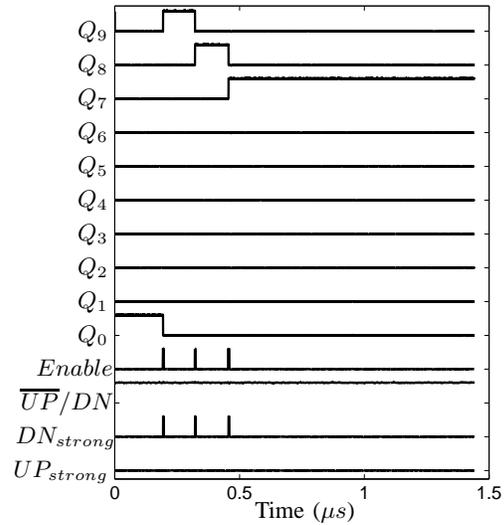}
\label{fig:Qs}}
\caption{Signals of the synchronizer obtained from 
layout extracted simulations.}
\label{fig:results}
\end{figure*}

\begin{figure}[h]
\centering
\psfrag{Timemicros}{\small{Two unit intervals}}
\psfrag{Voltage}{\small{data i/p (V)}}
\psfrag{datainput}{\small{Clock (V)}}

\includegraphics[width=6.5cm]{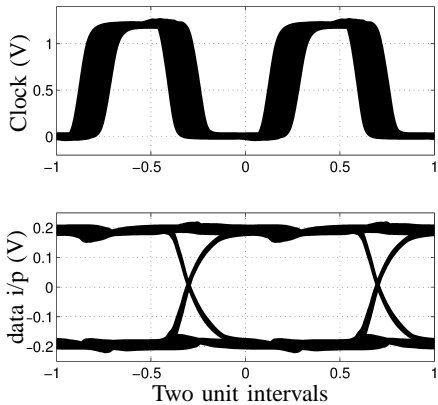}
\caption{Eye diagram of the data and the recovered clock 
obtained from layout extracted simuations. The jitter in the recovered clock 
is mainly due to the 5\% noise added to the supply for this simulation. 
Technology: \mbox{CMOS 130 nm}, 
Data rate: \mbox{2.5 Gbps}, \mbox{$V_{DD}$ = 1.2 V}.}
\label{fig:eye}
\end{figure}

\begin{figure}[h!]
\psfrag{UP}{\small{$UP_{strong}$}}
\psfrag{DN}{\small{$DN_{strong}$}}
\psfrag{C}{\tiny{C}}
\psfrag{D}{\tiny{D}}
\psfrag{Q}{\tiny{Q}}
\psfrag{RST}{\tiny{RST}}
\psfrag{Clock\_gate}{\small{$Enable$}}
\psfrag{Clock}{\small{Clock}}
\psfrag{UPDN}{\small{$\overline{UP}/DN$}}
\psfrag{Vc}{\small{$V_c$}}
\psfrag{0.75}{\small{$V_H$}}
\psfrag{0.25}{\small{$V_L$}}
\centering
\includegraphics[width=8.5cm]{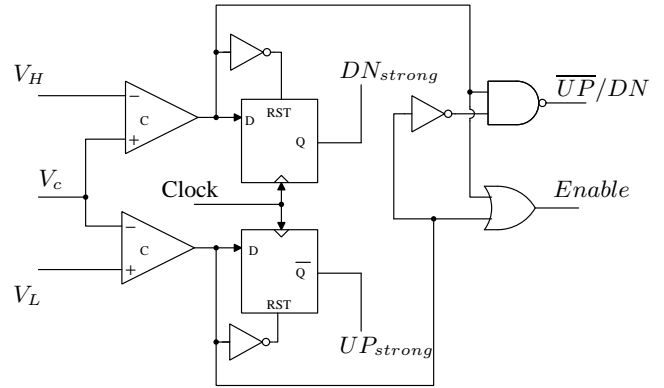}
\caption{Control logic for generating $\overline{UP}/DN$
and $Enable$ signals for the ring counter and $UP_{strong}$ \& $DN_{strong}$ signals
for the strong charge pump. $V_H$, $V_L$ are the 
upper and lower thresholds of window comparator respectively.}
\label{fig:logic}
\end{figure}

Fig. \ref{fig:results}\subref{fig:vc} shows the trajectory of the control voltage
from start-up to lock condition.
Fig. \ref{fig:results}\subref{fig:Qs} shows the progression of the ring counter
from reset state to locked state.
Fig. \ref{fig:eye} shows the simulated eye diagram
of the data and the clock after lock has been achieved.
Fig. \ref{fig:logic} shows the diagram of the window comparator
and the logic which controls the strong charge pump.
The comparators used in this circuit are traditional 
static comparators \cite{baker_cmos}.
The comparators trip when the control voltage crosses
the preset thresholds.
The clock for this circuit is obtained by dividing the system clock 
by a factor $K$.
The clock is divided to make the speed acceptable for the 
digital logic and the strong charge pump, in addition to saving power.
The ring counter is enabled only if any one of the comparator's outputs
is high. This happens only when the control voltage is not within the allowed
range. The $\overline{UP}/DN$ signal for the ring counter, and 
$UP_{strong}$ \& $DN_{strong}$ signals for the strong 
charge pump are derived from the 
comparator outputs. When $V_c$ is more than 
the upper threshold ($V_H$),
$\overline{UP}/DN$ is high. At this time the upper flip-flop is enabled.
On the next clock after this event, the ring counter counts down and in
order to bring $V_c$ within the window, the $DN_{strong}$ signal for the
strong charge pump is asserted. Once the loop filter capacitor discharges 
sufficiently so that $V_c$ is within the window comparator
thresholds, the comparator outputs go low. This resets the
flip-flops and de-asserts $Enable$. 
Depending on the chosen clock division ratio $K$,
the strong charge pump is designed
such that this exercise is completed in one cycle,
taking into consideration the time taken by the comparators to resolve.
Similar process follows
when $V_c$ is less than the lower threshold ($V_L$),
when the ring counter counts up.
The ring counter state can be
captured in a snapshot register and used to initialize
the counter for locking the loop quickly at subsequent power up.

\section{Clock domain transfer}
\label{sec:cdt}
\begin{figure}[h!]
\psfrag{Data}{\small{Data}}
\psfrag{Input}{\small{Input}}
\psfrag{phi\_s}{\small{$\phi_d$}}
\psfrag{phi\_i}{\small{$\phi_i$}}
\psfrag{RX Clock}{\small{$\phi_{Rx}$}}
\psfrag{DFF2}{\small{$\text{DFF}_{\text{i}}$}}
\psfrag{DFFRX}{\small{$\text{DFF\_R}_{\text{x}}$}}
\psfrag{Switch Matrix}{\small{Switch Matrix}}
\psfrag{Phase}{\small{Phase}}
\psfrag{Detector}{\small{Detector}}
\psfrag{feedback}{\small{Feedback}}
\psfrag{circuit}{\small{circuit}}
\psfrag{DLL}{\small{DLL}}
\psfrag{Clock}{\small{Clock}}
\psfrag{Domain}{\small{Domain}}
\psfrag{Transfer}{\small{Transfer}}
\psfrag{Receiver}{\small{Receiver}}
\psfrag{Error}{\small{Error}}
\psfrag{Retimed Data}{\small{Retimed Data}}

\centering
\includegraphics[width=8cm]{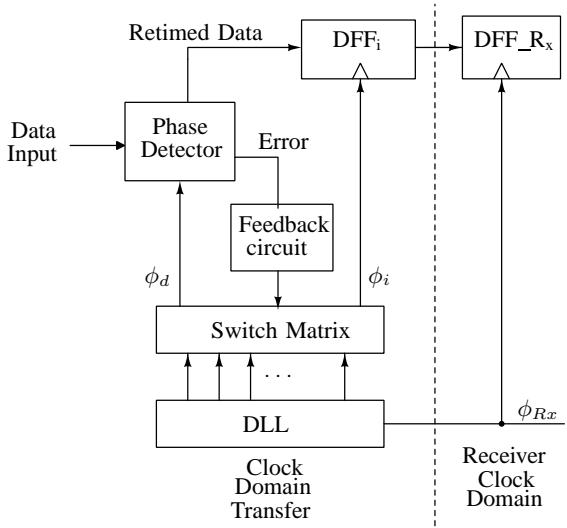}
\caption{Clock domain transfer from sampling clock $\phi_d$ 
to receiver clock $\phi_{Rx}$ (same as $\phi_0$ of the DLL). 
$\phi _i$ - clock of 
intermediate phase between $\phi_d$ and $\phi_{Rx}$.}
\label{fig:CDT}
\end{figure}

\begin{figure}[h!]
\centering
\psfrag{DFF1}{\small{$\phi_{d}$}}
\psfrag{DFF2}{\small{$\phi_{i}$}}
\psfrag{Rx}{\small{$\phi_{Rx}$}}
\psfrag{DLL}{\small{DLL phases}}
\psfrag{phi0}{\small{$\phi_{0}$(=$\phi_{Rx}$)}}
\psfrag{phiN}{\small{$\phi_{7}$}}
\psfrag{dots}{\ldots}
\includegraphics[width=8cm]{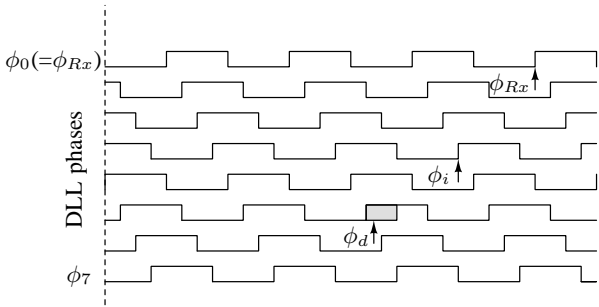}
\caption{Illustration of the timing diagram of clock phase recovery system
with clock domain transfer, under the worst case
latency.}
\label{fig:cdt_ck_an}
\end{figure}

Once the data has been sampled with a synchronized clock
and converted to CMOS levels, it should be re-timed
to the receiver clock domain.
Typically, multi flip-flop serial synchronizers are used for this. 
Since multiple clock phases are already available in the design described here,
the clock domain transfer can be performed with a single flip-flop
that is clocked with an intermediate phase.
Fig. \ref{fig:CDT} shows the serial synchronizer along with the 
clock domain transfer flip-flop.
Here the DLL generates
$N$ phases of the clock. The clock recovery system
selects the phase closest to the center of the data
($\phi_n$) and delays it appropriately, to generate
$\phi_d$ such that it
is positioned at the center of the eye. The retimed
data from the phase detector is available at this
phase $\phi_d$. The receiver
samples the data at the receiver clock which is
$\phi_{Rx}$ (which is same as $\phi_0$ of the DLL). 
The flip-flop $\text{DFF}_{\text{i}}$ should be clocked
with a phase that will guarantee that no flip-flop
samples during a data transition. $\phi _i$ is
selected as
\begin{align*}
\phi _i &= \phi _{(n+2 - \frac{N}{2})} \hspace{1cm} &\mbox{if } n+2 > \frac{N}{2}
\\
        &= \phi _0 \hspace{1cm} &\mbox{otherwise.} 
\end{align*}
Fig. \ref{fig:cdt_ck_an}
shows the sampling clocks for the worst case latency, for an example
case that uses an 8 phase DLL ($N=8$).
One can see that $\phi _{(n+2 - \frac{N}{2})} \sim \overline{\phi _0}$.
Using $\overline{\phi _0}$ simplifies the implementation in two ways.
First, it reduces the load on the DLL and switch matrix. In addition, it also 
compensates for the delay introduced by the clock buffers from switch 
matrix output to the phase detector's clock input, with the delay 
introduced in generating $\overline{\phi _0}$ from $\phi _0$. 

Here, an assumption that a flip-flop sampling a valid
data input is able to resolve its output in time
$\frac{T}{2}-T_{setup}$ is made. Since pipelined
microprocessors generally have a logic depth of more than 
5 NAND gates, this assumption is not very demanding.
The reason for choosing $\phi _i = \phi _{(n+2 - \frac{N}{2})}$
for $n+2 > \frac{N}{2}$ comes from the fact that the
fine tuning loop of the clock recovery
system could delay the sampling clock by at most
2 phase steps of the DLL. This is because the VCDL 
is overdesigned for two phase steps so as to meet the 
range requirements across process corners.
This is explained later in section \ref{sec:implementation} when discussing the VCDL.
The total latency of the above synchronizer is
at most 3 clock cycles. The phase detector takes 2
clock cycles (as will be explained in section \ref{sec:implementation})
 and the following DFF's
output is sampled half a clock cycle later.
Under the worst conditions the third flip-flop 
in the chain will introduce another
half a clock cycle delay. This makes the
total latency $\leq 3T$.

\section{Jitter Analysis}
\label{sec:jitter}

\begin{figure}[h!]
\centering
\psfrag{Timemicros}{\small{Time ($\mu s$)}}
\psfrag{Controlvoltage}{\small{$V_c$ (V)}}
\psfrag{datainput}{\small{Data i/p (V)}}
\psfrag{Twounitintervals}{\small{Two unit intervals}}
\includegraphics[width=8cm]{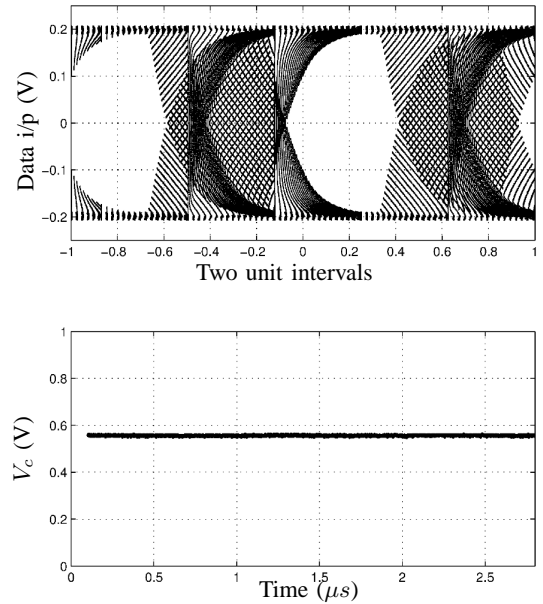}
\caption{Simulated eye diagram and $V_c$ with a sinusoidal jitter of frequency 
1 MHz and amplitude of 0.5 UI. The transmitter and receiver clock 
jitter are correlated.}
\label{fig:data_1Mjitter}
\end{figure}

Clock recovery circuits that use the data transitions to estimate 
the correct sampling phase are generally sensitive to the jitter
in the received data. The jitter in the received data comes due to
the phase noise in the transmitter clocks which is random in nature,
and due to ISI in the channel, which is deterministic.
Also, wander in the clock generating
oscillators results in low frequency jitter. Typically, clock 
recovery circuits are required to tolerate small amplitude of 
high frequency jitter (typically under 0.1 unit interval (UI)) and large
amplitude of low frequency jitter (up to 0.5 UI or more). 
However, the synchronizer for the application of repeaterless on-chip interconnect 
system discussed in this paper, does not need to be tolerant to the low
frequency jitter. This is because the transmitter and the receiver share the same 
clock source albeit in arbitrary phase relationship. To verify this
the synchronizer was tested with a low frequency jitter of \mbox{1 MHz}
frequency and 0.5 UI amplitude. In order to reduce the simulation 
time, first, extracted layout of the DLL alone was simulated with
an input clock having low frequency jitter, which confirmed that the DLL clock
phases tracked the low frequency jitter. 
Then the full loop of the synchronizer with the schematic level
netlists and an ideal DLL model was simulated.
Fig. \ref{fig:data_1Mjitter} shows the eye diagram of the 
received data with jitter, and the control voltage $V_c$ for this simulation. 
As the low frequency jitter is correlated between 
the transmitter and the receiver, the control 
voltage does not have to track it.
\begin{figure}[h!]
\centering
\psfrag{Timemicros}{\small{Time ($\mu s$)}}
\psfrag{Controlvoltage}{\small{$V_c$ (V)}}
\psfrag{datainput}{\small{Data i/p (V)}}
\psfrag{Twounitintervals}{\small{Two unit intervals}}
\includegraphics[width=8cm]{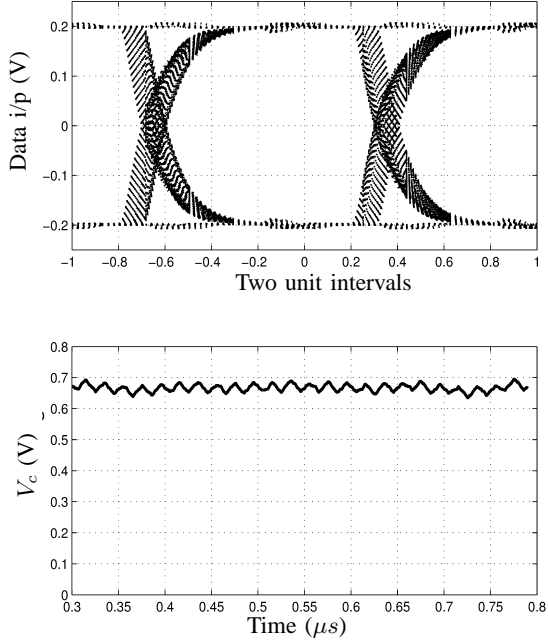}
\caption{Simulated eye diagram and $V_c$ with a sinusoidal jitter of frequency
50 MHz and amplitude of 0.1 UI. Jitter is added only to transmitter clock.}
\label{fig:data_50Mjitter}
\end{figure}

High frequency jitter which is generated in the clock distribution 
networks due to thermal noise in transistors and power supply noise
will however not be common to the transmitter and receiver. 
This high frequency jitter is however filtered out by the filter
capacitor in the synchronizer circuit. Fig. \ref{fig:data_50Mjitter} shows
the eye diagram and control voltage for a sinusoidal jitter of an amplitude 
of 0.1 UI at a frequency of 50 MHz. 
The circuit was also simulated for
high frequency uncorrelated jitter of \mbox{50 MHz} and 200 MHz 
for 10 $\mu s$ with various jitter amplitudes. 
The circuit has no errors even for jitter amplitudes 
as high as 0.4 UI.

In conclusion for an 
acceptable level of low frequency jitter tolerance, the DLL in the 
synchronizer must be designed with a loop bandwidth that is 
more as compared to the expected low frequency jitter 
in the clock generating PLL.

\section{Implementation details and Results}
\label{sec:implementation}
\subsection{Implementation Details}
\begin{figure}[h!]
\psfrag{Vc__}{\small{$V_{in}$}}
\psfrag{VH__}{\small{$V_H$}}
\psfrag{Vout_}{\small{$V_o$}}
\psfrag{ns}{\footnotesize{6 ns}}
\psfrag{Timemicros}{\small{Time ($\mu s$)}}
\psfrag{VoltageV}{\small{Voltage (V)}}
\centering
\includegraphics[width=6.3cm]{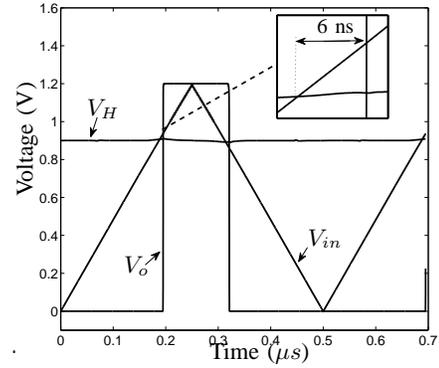}
\caption{Simulation of the static comparators with an input
slope generated by the weak charge pump driving the loop filter capacitor.}
\label{fig:static_comp}
\end{figure}

The clock synchronizer has been designed in \mbox{130 nm} UMC MM CMOS technology,
with a supply voltage of 1.2 V.
The window comparator thresholds in the coarse tuning loop
were $V_{DD}/4$ and $3V_{DD}/4$. The comparator is tested for an input 
slope of 1 $\mu$A/200 fF.
Fig. \ref{fig:static_comp} shows the simulated response of the comparator 
under these test conditions. The comparators
resolve in about 6 ns after the input crosses the threshold.
A clock division ratio ($K$) of 16 
was used for the coarse tuning loop.

\begin{figure}[h!]
\psfrag{UP}{\small{UP}}
\psfrag{DN}{\small{DN}}
\psfrag{D}{\small{D}}
\psfrag{Q}{\small{Q}}
\psfrag{C}{\small{C}}
\psfrag{Data}{\small{Data}}
\psfrag{Clock}{\small{Clock}}
\centering
\includegraphics[width=8cm]{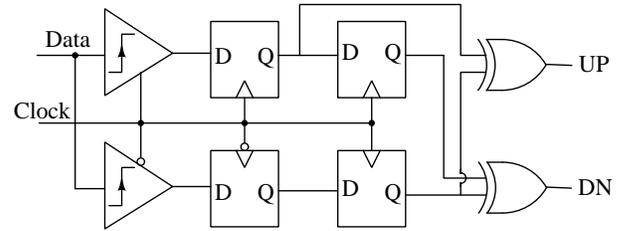}
\caption{Circuit diagram of the Alexander Bang Bang phase detector.}
\label{fig:alex}
\end{figure}

The Alexander bang bang phase detector was used as the phase detector. 
Fig. \ref{fig:alex} shows the diagram of the phase detector
used in the design.
Sense-amplifier based clocked comparators \cite{Nikolic-jssc-2000}
were used and were followed with another flip-flop 
clocked by the same clock.
This is done because the sense-amplifier comparator can take
more than half a clock cycle to resolve,
and for proper generation of the $UP$ and $DN$ pulses
the comparator is required to resolve in less than half a clock cycle.
It is interesting to note that in the case
when the Alexander phase detector is used
to recover only the phase of received clock from the data,
the loop can sometimes remain stuck
with the wrong edge at the center of the eye diagram.
This is a rare phenomenon and occurs only
when certain conditions are met.
This effect is discussed in Appendix I.

\begin{figure}[h]
\psfrag{UP}{\small{UP}}
\psfrag{DN}{\small{DN}}
\psfrag{DNB}{\small{$\overline{\text{DN}}$}}
\psfrag{UPB}{\small{$\overline{\text{UP}}$}}
\psfrag{1u}{\small{$1\mu A$}}
\psfrag{M1}{\small{$M_1$}}
\psfrag{M2}{\small{$M_2$}}
\psfrag{M3}{\small{$M_3$}}
\psfrag{M4}{\small{$M_4$}}
\psfrag{M5}{\small{$M_5$}}
\psfrag{M6}{\small{$M_6$}}
\psfrag{M7}{\small{$M_7$}}
\psfrag{M8}{\small{$M_8$}}
\psfrag{M9}{\small{$M_9$}}
\psfrag{M10}{\small{$M_{10}$}}
\psfrag{M11}{\small{$M_{11}$}}
\psfrag{M12}{\small{$M_{12}$}}
\psfrag{M13}{\small{$M_{13}$}}
\psfrag{M14}{\small{$M_{14}$}}
\psfrag{M15}{\small{$M_{15}$}}
\psfrag{UPS}{\small{UP$_{\text{strong}}$}}
\psfrag{DNS}{\small{DN$_{\text{strong}}$}}
\psfrag{Vc}{\small{V$_\text{c}$}}
\psfrag{200}{\small{200fF}}
\centering
\includegraphics[width=8cm]{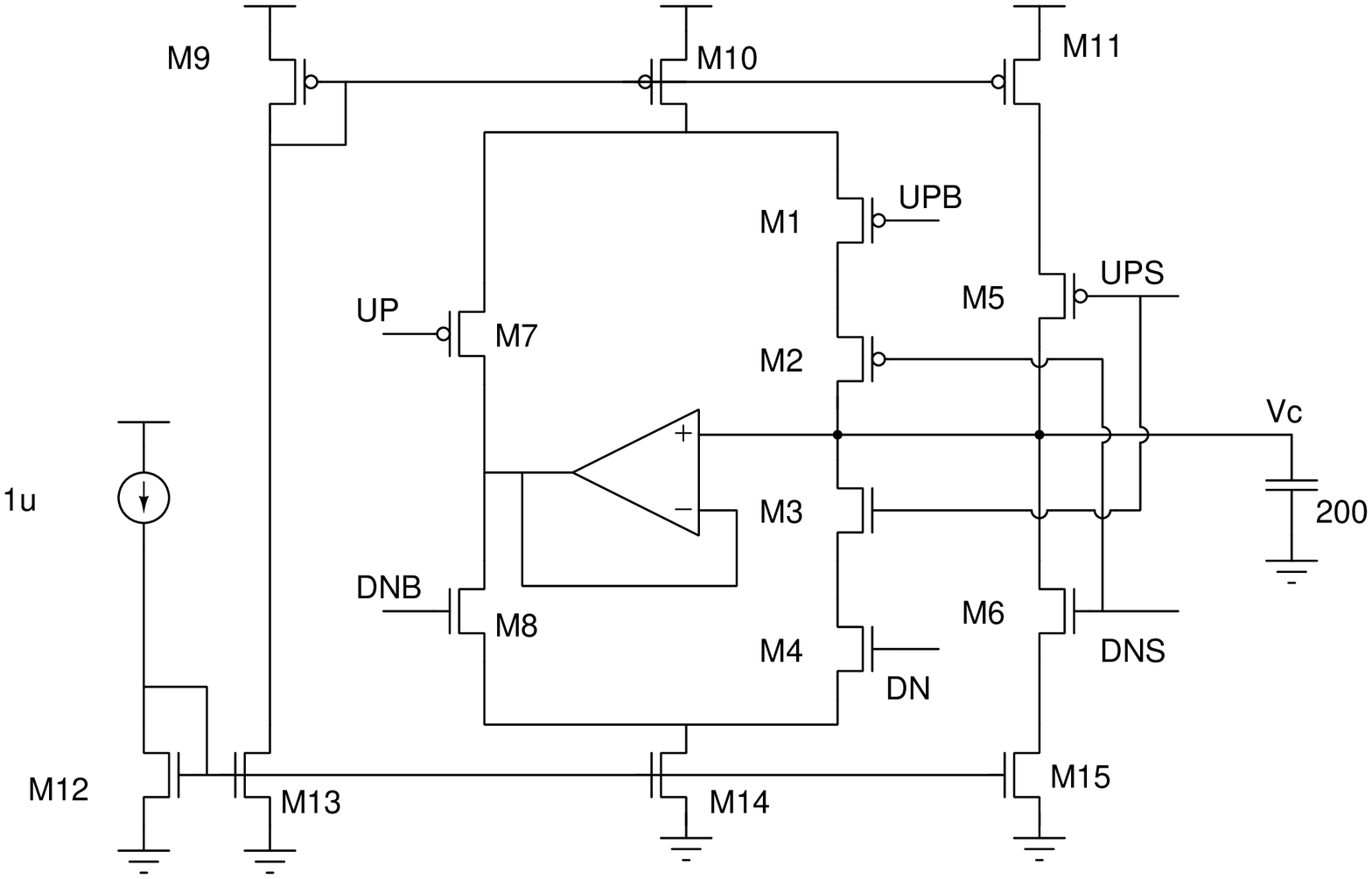}
\caption{The weak and the strong charge pump. DN, $\overline{\text{DN}}$,
UP and $\overline{\text{UP}}$ are from the fine tuning loop and UP$_{\text{strong}}$
and DN$_{\text{strong}}$ are from the coarse tuning loop.}
\label{fig:charge_pump}
\end{figure}

Fig. \ref{fig:charge_pump} shows the circuit 
diagram of the charge pump,
which is the well known active amplifier charge pump circuit 
\cite{active_CP}, modified to add the strong current source and sink, which
injects current into the loop filter capacitor along with
the weak current sources.
$DN$, $\overline{DN}$, $UP$ and $\overline{UP}$
are driven by the fine tuning loop
and $UP_{strong}$ and $DN_{strong}$ are driven
by the coarse tuning loop.
The strong charge pump is designed to be
16 times the strength of the weak charge pump.
The weak charge pump is of 1$\mu$A.
The loop filter capacitor is a MIMCAP of \mbox{200 fF}. 
The geometries of the transistors in the
charge pump are as follows.
%given in Table \ref{tbl:cp-sizes}.
%
%\begin{table}
\begin{center}
%\centering
\begin{tabular}{|c|c|c|}
\hline
Transistor & W & L \\ \hline
$M_1,M_2$ & 950 nm & 120 nm \\ \hline
$M_3,M_4$ & 320 nm & 120 nm \\ \hline
$M_5$ & 1 $\mu$m & 120 nm \\ \hline 
$M_6$ & 300 nm & 120 nm \\ \hline
$M_7$ & 700 nm & 120 nm \\ \hline
$M_8$ & 160 nm & 120 nm \\ \hline
$M_9,M_{10}$ & 2 $\mu$m & 500 nm \\ \hline
$M_{11}$ & 17 $\mu$m & 500 nm \\ \hline
$M_{12},M_{13},M_{14}$ & 500 nm & 500 nm \\ \hline
$M_{15}$ & 7 $\mu$m & 500 nm \\ \hline 
\end{tabular}
\end{center}
%\caption{Transistor sizes of the charge pump circuit \mbox {in Fig. \ref{fig:charge_pump}}}
%\label{tbl:cp-sizes}
%\end{table}

%
%
%$M_1,M_2=950nm/120nm$, $M_3,M_4=320nm/120nm$, \\
%$M_5 = 1\mu m/120nm$, $M_6 = 300nm/120nm$, \\
%$M_7 = 700nm/120nm$, $M_8 = 160nm/120nm$, \\
%$M_9,M_{10} = 2\mu m/500nm$, $M_{11} = 17\mu m/500nm$, \\
%$M_{12},M_{13},M_{14} = 500nm/500nm$,
%$M_{15} = 7\mu m/500nm$.

The extra transistors $M_2$ and $M_3$ are used to disable the 
weak charge pump when the strong charge pump is active. This is 
done so as to prevent the weak current source (sink)  and 
strong current sink (source) from forming a path
from VDD to GND, which would push the transistors to their linear region. 
The opamp used is a traditional single stage differential amplifier.

\begin{figure}[h!]
\psfrag{IN}{\small{IN}}
\psfrag{OUT}{\small{OUT}}
\psfrag{VCN}{\small{$V_{cn}$}}
\psfrag{VCP}{\small{$V_{c}$}}
\centering
\includegraphics[width=8cm]{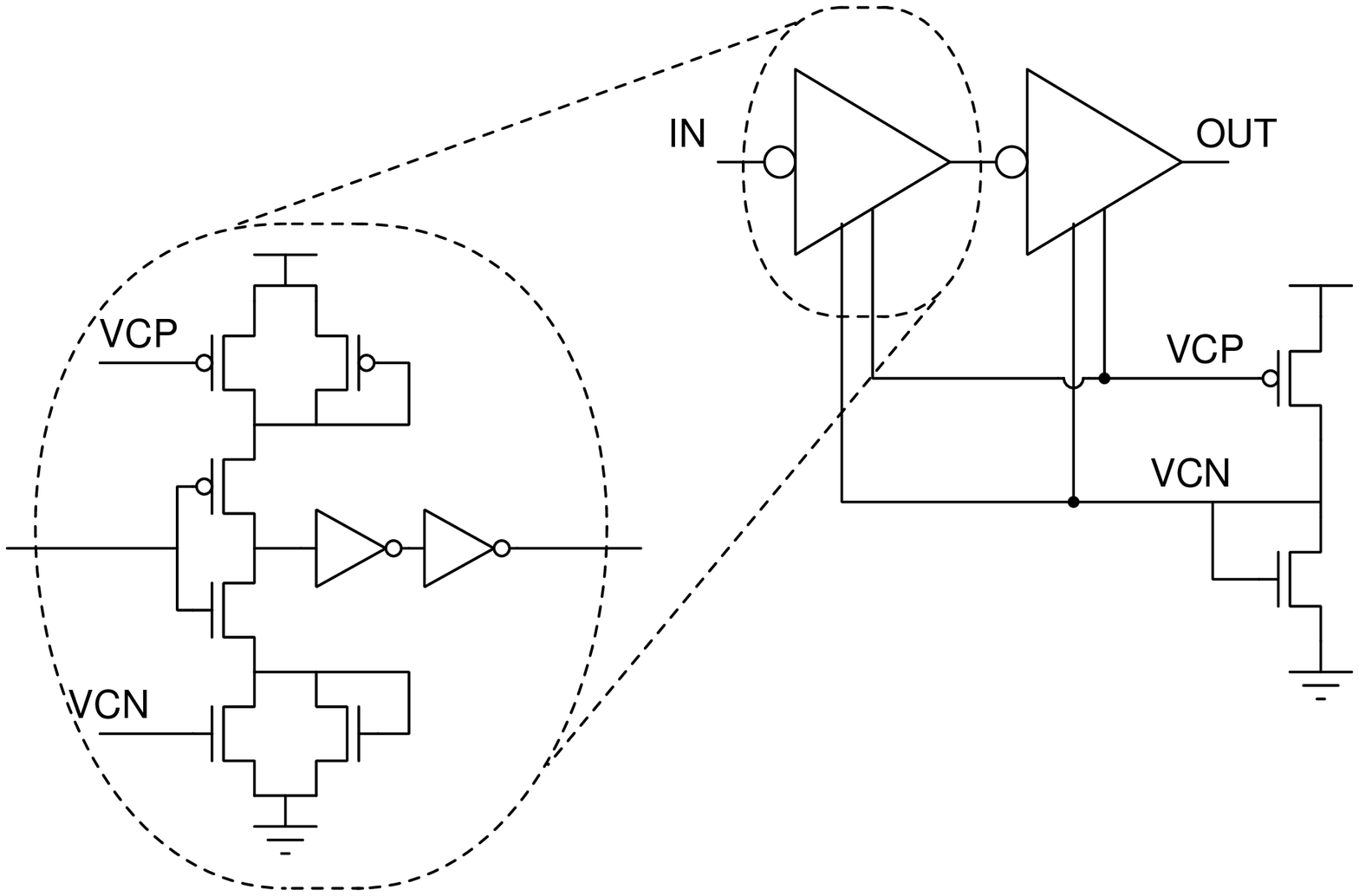}
\caption{Circuit schematic of the voltage controlled delay line (VCDL).}
\label{fig:vcdl}
\end{figure}

\begin{figure}[h!]
\centering
\psfrag{SS}{\small{SS}}
\psfrag{TTTT}{\small{TT}}
\psfrag{FFFF}{\small{FF}}
\psfrag{FNSPPPP}{\small{FNSP}}
\psfrag{SNFP}{\small{SNFP}}
\psfrag{controlvoltage}{\small{\hspace{-2.5em}Control Voltage $V_c$ (V)}}
\psfrag{Normalizeddelay}{\small{\hspace{-1em}Normalized Delay ($ps$)}}
\includegraphics[width=8cm]{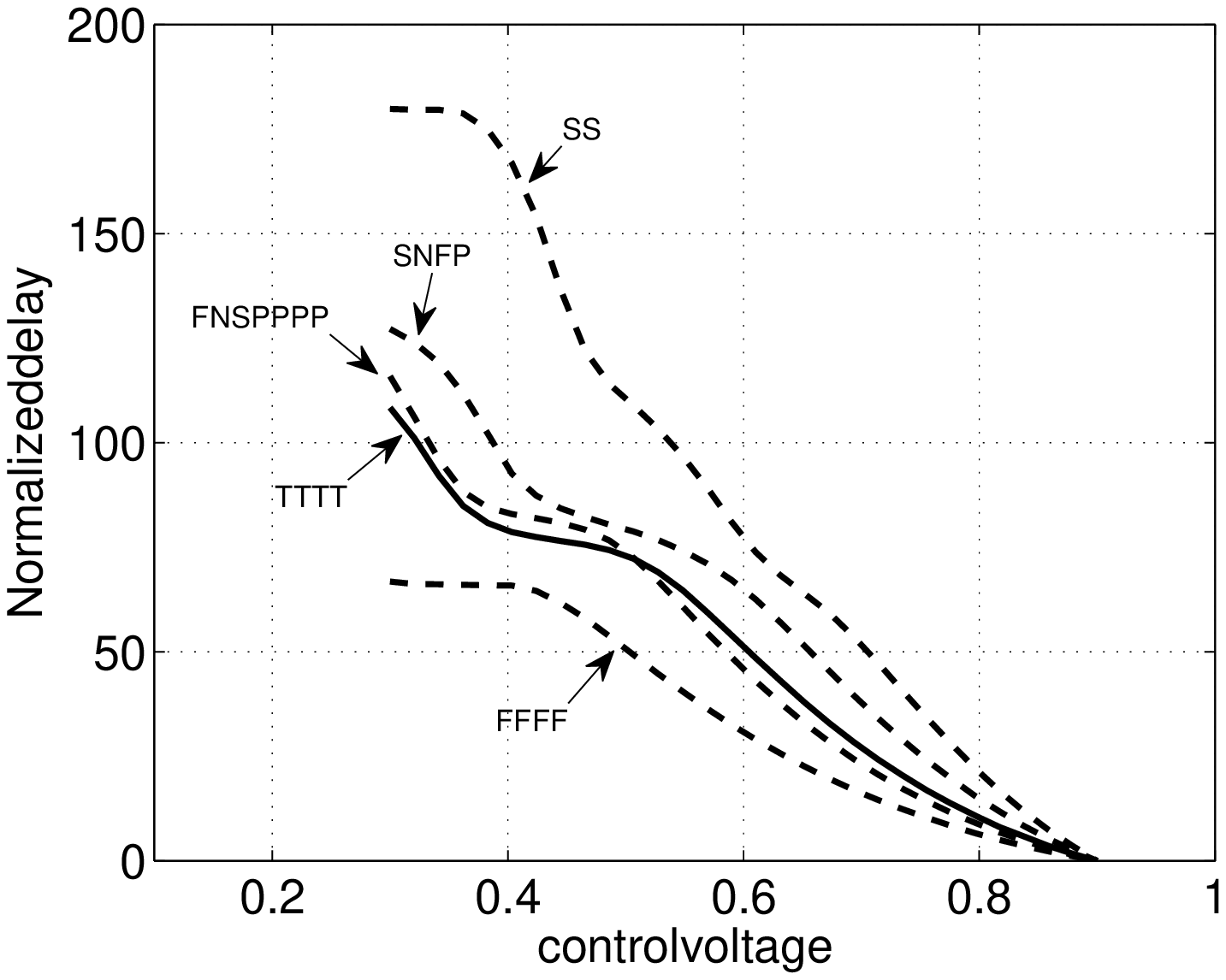}
\caption{Transfer characteristics of the VCDL across process corners.}
\label{fig:vcdl_range}
\end{figure}

Fig. \ref{fig:vcdl} shows the implementation of the VCDL
of the fine tuning loop. The delay cells are current starved inverters,
with the current sources in parallel with diode connected transistors which
act as bleeder resistors as shown. Two stages are used in cascade to get the
required tuning range. 
Fig. \ref{fig:vcdl_range} shows the range of delays for the allowed
range of control voltages across different process corners.
\begin{figure}[h]
\centering
\psfrag{L}{\small{76$\mu m$}}
\psfrag{W}{\small{80$\mu m$}}
\psfrag{Vc}{\small{$V_c$}}
\includegraphics[width=6.5cm]{./chip}
\caption{Photograph of a die of the fabricated circuit.}
\label{fig:chip}
\end{figure}
The system does not impose any linearity requirement on the VCDL. However,
the transfer characteristics must be monotonic. The VCDL must be designed 
to have a range of greater than 1 phase step of the DLL.
In order to meet this requirement the VCDL must be overdesiged. 
When designed for a range of 1 phase step in the fastest corner,
the VCDL has a range of 2 phase steps under typical process parameters.
A DLL of 10 phases was implemented. Linear delay cells
reported in \cite{delay_cell_2008} are used, with a precharge phase 
detector. The complete synchronizer has an area of 
\mbox{76$\mu$m $\times$ 80$\mu$m}. 
Fig. \ref{fig:chip} shows a photograph of
a bare die of the fabricated chip.

\subsection{Results and Comparisons}
The control voltage $V_c$ of the synchronizer and the
control voltage of the DLL that generates the multiple phases of the clock,
were buffered with internal opamps and brought out on pins for
testing. The status of the ring counter in the receiver is
also encoded and brought out on pins. The transmitter in the
test circuit was a 15 bit PRBS with a low swing transmitter.
A short interconnect is used between the transmitter and receiver. For
testing the receiver's phase tracking, the transmitter's clock 
is deliberately shifted using an
inverter based delay line, with a programmable tap.
%
%could be shifted
%for artificially tuning its phase relative to the receiver clock
%using an inverter based delay line, whose tap was externally
%programmable. 
 
%Figure \ref{fig:pcb} shows a photograph of the test board with the
%packaged chip.
%

%\begin{figure}
%\centering
%\includegraphics[width=7cm]{./figs/pcb}
%\caption{Photograph of the test board.}
%\label{fig:pcb}
%\end{figure}

\begin{figure}[h!]
\centering
\includegraphics[width=8cm]{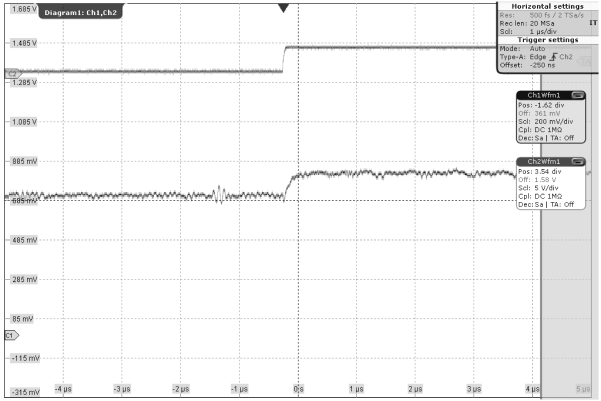}
\caption{Measured trajectory of $V_c$ with an introduced
 phase shift in the transmitter clock that 
does not need a DLL increment/decrement. The upper waveform is the
transmitter phase shift clock.}
\label{fig:nojump}
\end{figure}

\begin{figure}[h!]
\centering
\includegraphics[width=8cm]{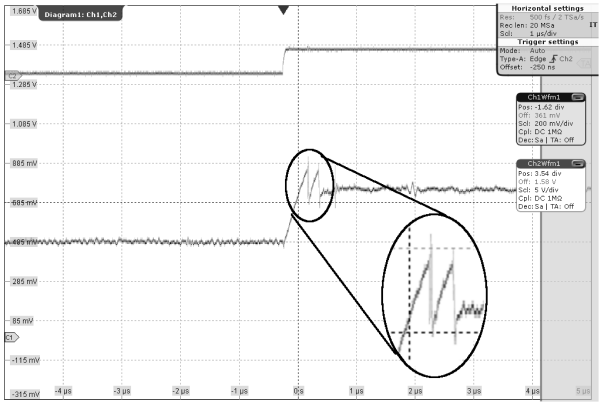}
\caption{Measured trajectory of $V_c$ with an introduced
phase shift in the transmitter clock that 
needs DLL phase decrements. The upper waveform is the transmitter
phase shift clock.}
\label{fig:withjump}
\end{figure}

The fabricated chip was tested at a frequency of 1.3 GHz, and
the power consumption of the complete synchronizer was \mbox{1.4 mW}
off a 1.2V supply. 
The chip was found to lie in the FNSP (fast N, slow P) corner.
The clock was generated using a three stage inverter ring oscillator,
whose frequency was controlled by modulating its supply.
No duty cycle correction circuit was included.
The tracking of the phase detector
was confirmed by deliberately shifiting the clock
phase of the transmitter and observing the control voltage 
of the circuit.
%
%introducing artificial phase shifts between 
%the transmitter and receiver, and observing the analog control voltage
%and the counter state. 
Fig. \ref{fig:nojump}
shows the trajectory of the control voltage ($V_c$)
when the introduced phase shift in the transmitter does not need
a DLL phase increment or decrement. Fig. \ref{fig:withjump} shows the 
trajectory of the control voltage when the introduced phase shift in the
transmitter needs DLL phase decrements for achieving lock. Once 
lock is achieved, it was observed that the circuit remains locked
over long periods of time, 
which was tested by monitoring the lock over 
a period of up to 30 minutes.

\begin{figure}[h!]
\centering
\includegraphics[width=8cm]{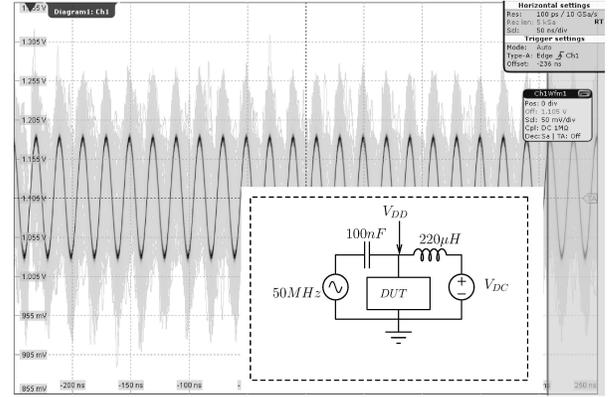}
\caption{Waveform showing the supply voltage modulated with a $50 MHz$, 
$\pm 80mV$ sine wave. Inset is the circuit used to generate the 
modulated supply.}
\label{fig:supply_nw}
\end{figure}

\begin{figure}[h!]
\centering
\includegraphics[width=8cm]{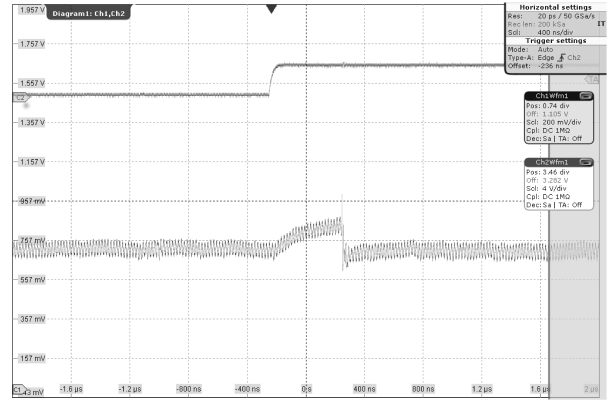}
\caption{Measured trajectory of $V_c$ with supply modulated with
a $50MHz$, $\pm 80mV$ sine wave. The upper waveform is the 
transmitter phase shift clock.}
\label{fig:withnoise}
\end{figure}

The circuit was also tested with supply voltage fluctuations 
by adding a $50 MHz$ $\pm 80mV$ sine wave to the supply voltage.
Fig. \ref{fig:supply_nw} shows the supply voltage as observed on an
oscilloscope along with an inset of the circuit used to add a 
sine wave over the DC supply. 
Fig. \ref{fig:withnoise} shows the trajectory of the control voltage for 
the measurement under these conditions.

To confirm scalability of the architecture,
the synchronizer was also designed and simulated in TSMC 65 nm CMOS technology,
for a data rate of \mbox{4 Gbps}.
The clock division ratio ($K$) of 32 was used for
deriving the clock for the coarse tuning loop. Rest of the implementation details
are identical to the 130 nm implementation.
From layout extracted simulations, the power consumption was found to be
\mbox{1.5 mW} drawn from a 1 V supply. The area of the synchronizer is 
\mbox{$48\mu m\times 50\mu m$}.

Table \ref{tbl:compare} compares the performance of 
previously reported clock synchronizers for repeaterless interconnects
with the presented design.
As seen from Table \ref{tbl:compare}, the proposed 
synchronizer for repeaterless low swing interconnects 
is the only one for on-chip interconnects
that is adaptive and does not have phase quantization error. 
Also, to the best of the authors knowledge, 
this is the only work which discusses clock domain transfer to receiver
clock domain for low swing interconnects.

\begin{table*}
\begin{center}
\caption{Comparison with other reported repeaterless synchronizers}
\label{tbl:compare}
\begin{tabular}{|c|c|c|c||c|c|}
\hline 
 & JSSC'14 \cite{Lee-jssc14} & JSSC'05 
\cite{phase_interpolator} & JSSC'03 \cite{dig_pi_jssc03} & 
\multicolumn{2}{c|}{This work} \\ \hline 
Process & 65 nm & 110 nm & 110 nm & 130 nm & 65 nm \\
 & & & & (Measured) & (Simulated) \\ \hline
Interconnect type & On-chip & Off-chip & Off-chip & \multicolumn{2}{c|}{On-Chip}\\ \hline
Synchronizer type & Mesochronous & Plesiochronous & Plesiochronous & \multicolumn{2}{c|}{Mesochronous} \\ \hline
Controller type & Digital & Analog & Digital & \multicolumn{2}{c|}{Coarse digital + fine analog} \\ \hline
Adaptive control & No & Yes & Yes & \multicolumn{2}{c|}{Yes} \\ \hline
Clock domain transfer & No & No & No & \multicolumn{2}{c|}{Yes} \\ \hline
Data rate & 3 Gb/s & 10 Gb/s & 10 Gb/s & 1.3 Gb/s & 4 Gb/s \\ \hline
Power consumption & 1.08 mW & 220 mW & 129 mW & 1.4 mW & 1.5 mW  \\ \hline
Supply Voltage & 0.9 V & 1.5 V & 1.2 V & 1.2 V & 1 V \\ \hline
Area ($\mu m\times \mu m$) & Not reported & 250 $\times$ 1400 & 1600 $\times$ 2600 & 76 $\times$ 80 & 48 $\times$ 50\\ \hline
\end{tabular}
\end{center}
\end{table*}

\section{Conclusion}
\label{sec:conclude}
This paper presents a clock synchronizing circuit for 
repeaterless low swing on-chip interconnect. The circuit uses a 
coarse and a fine correction loop, and permits the lock state of the
coarse correction loop to be saved and recalled on subsequent power up, 
for quick locking. A low latency clock domain transfer then
transfers the data to the receiver clock domain. The circuits have been designed
and tested in CMOS 130 nm technology. Further, to verify the scalability 
the circuit, the circuit was also designed and simulated in CMOS 65 nm technology.

\appendices
\section{False edge locking in DLL based CDR loops using the Alexander phase detector}
\label{app:APD}
\begin{figure}[htb!]
\centering
\includegraphics[width=7cm]{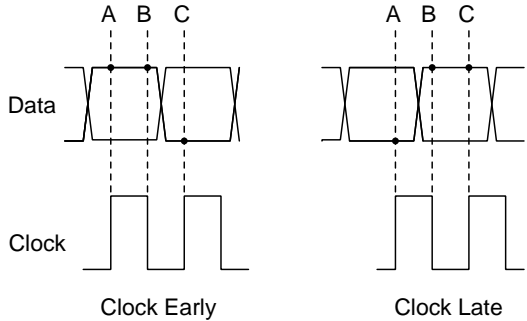}
\caption{The sampling instants of the Alexander phase detector.}
\label{fig:alex_sample}
\end{figure}

The Alexander phase detector is widely used for measuring the phase
difference between data and clock, as required in clock and data
recovery circuits \cite{CDR_razavi}. The circuit diagram of the
phase detector is shown in \mbox{Fig. \ref{fig:alex}}.
This phase detector samples the data at two points per bit period to
make a decision whether the clock is leading or lagging the data.
\mbox{Fig. \ref{fig:alex_sample}} shows the sampling instants
when the clock is late and early.

The UP and DN signals are then derived
from the three consecutive samples A, B and C as
\begin{align}
\text{DN} &= B \oplus C \nonumber \\ 
\text{UP} &= A \oplus B \nonumber
\end{align}

This evaluation is performed on the active edge of the clock,
and the last three consecutive samples are used.
The UP and DN signals are then integrated using a charge 
pump that controls the clock frequency/phase.
The negative feedback loop brings the clock to the 
center of the eye.

\begin{figure}[htb!]
\centering
\includegraphics[width=7cm]{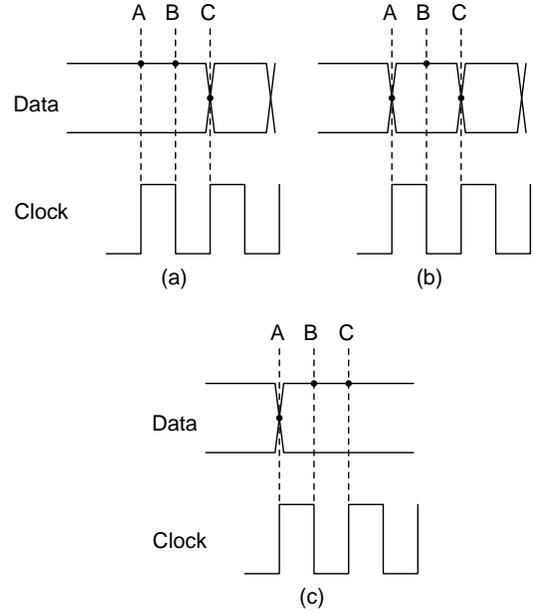}
\caption{The sampling instants of the Alexander phase detector, when 
the phase difference between data and clock in $\pi$ radians.}
\label{fig:Alex_sample_wrong}
\end{figure}

When certain, albeit unlikely, conditions are met the 
Alexander phase detector can lock to the
wrong edge at the center of the eye. This happens when 
the frequencies of the data and the clock 
are exactly equal, and the phase of the 
clock is exactly $\pi$ radians offset from
the center of the eye and the data has a 50\% activity.
Under such conditions the phase detector
can lock to the correct edge either by increasing 
the relative phase by $\pi$ radians,
or by reducing the relative phase by $\pi$ radians. Theoretically, the 
phase detector can take infinite time to choose one solution over 
the other \cite{buridan}.
Fig. \ref{fig:Alex_sample_wrong} shows the sampling instants
for various data transition permutations, when the data
and clock have a phase error of $\pi$ radians.
As seen in Fig. \ref{fig:Alex_sample_wrong} (a), if sample
C resolves to the same values as A and B no corrective action is performed.
If however, C is not equal to A and B, then the phase detector asserts
a DN signal. Similarly if this is followed with pattern shown in 
Fig. \ref{fig:Alex_sample_wrong} (c), only if A resolves unequal to 
B and C, the phase detector asserts an UP signal. Since the deciding samples
are taken within the metastability window of the comparators, these can
occur with equal probability nullifying each other. The case in 
Fig. \ref{fig:Alex_sample_wrong} (b) can also cause an additional
case where A and C resolve to a value different from sample at B.
This causes the phase detector to assert both UP and DN, which
results in no change on the control voltage. However, this case
was not observed in simulations, but it is possible in principle.
The above conditions can cause the net result such
that the phase detector is stuck at the wrong edge.

\begin{figure}[h!]
\centering
\psfrag{VC}{\small{$V_c$}}
\psfrag{Voltage}{\small{Voltage (V)}}
\psfrag{Timeseconds}{\small{Time ($\mu s$)}}
\includegraphics[width=8cm]{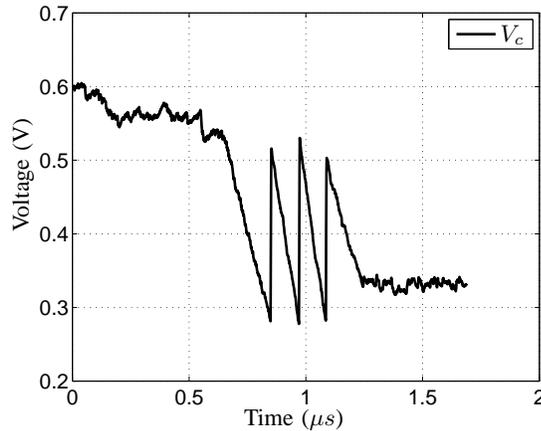}
\caption{Control voltage ($V_c$) of the phase detector when the phase detector is locked
to the wrong edge at the center of the eye.}
\label{fig:Alex_wrong_edge}
\end{figure}

Generally the randomness in the data and system noise is sufficient to 
get the phase detector out of this zone.
False locking happens when the phase detector's correct sampling edge is within the
meta-stability window of the comparator under the initial conditions. Since the
meta-stability window increases after layout, the inertia of staying locked at the false
edge is higher when layout extracted circuits are tested. Fig. \ref{fig:Alex_wrong_edge} shows
the control voltage ($V_c$) waveform when the phase detector is locked to the wrong edge. 

The
simulation was performed on pre-layout (schematic) circuit netlist. It is seen that
around \mbox{0.7 $\mu$s} simulation time, the randomness of the data pushes the clock edge out
of the unstable equilibrium zone of the loop and lock to correct edge is achieved in the subsequent
\mbox{0.6 $\mu$s.}

It may be noted that this phenomenon, in principle, can also occur when the phase detector
is used for frequency and phase locking. However, the sweeping of the clock edge across the
data eye due to the difference in the clock and data frequency; the required phase detector gain; and the required data sequence
to hold the phase detector at the wrong edge; make its probability of occurrence extremely remote.

This problem is not expected to be severely limiting, as noise will eventually
bring the system out of the unstable equilibrium. This problem if at all
occurs, will only occur the first time around. Once correct lock
is achieved, and the state is saved, this false lock will never occur,
as the recalled state brings the loop state very close to the correct lock.
By using phase detectors that sample the data at more than two times per bit
period \cite{oversampling_pd}, one can eliminate the probability of the first
time occurrence as well.

\bibliographystyle{IEEE}
\bibliography{IEEEabrv,cdr_ref-new}
%\bibliography{cdr_ref}

\begin{IEEEbiography}[{\includegraphics[width=1in,height=1.25in,clip,keepaspectratio]{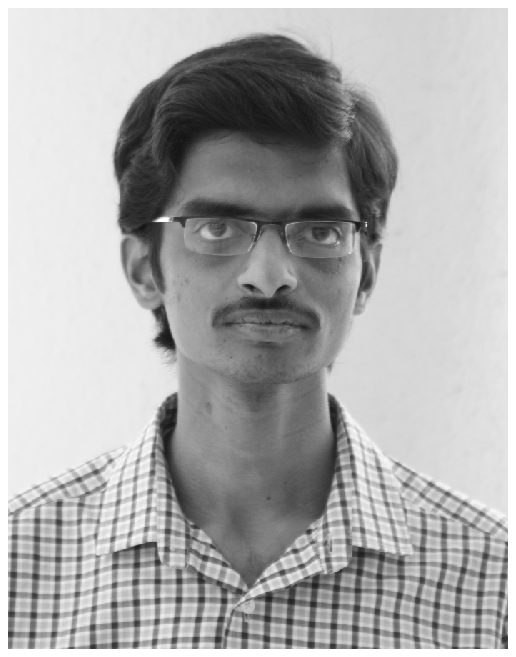}}]{Naveen Kadayinti}
%\begin{IEEEbiography}{Naveen Kadayinti}
is a research scholar at the Indian Institute of Technology
Bombay, and is currently working towards 
his thesis on ``High speed interconnects". 
His research interests include wired and wireless communication circuits
and mixed signal SoC design and test.
\end{IEEEbiography}
\begin{IEEEbiography}[{\includegraphics[width=1in,height=1.25in,clip,keepaspectratio]{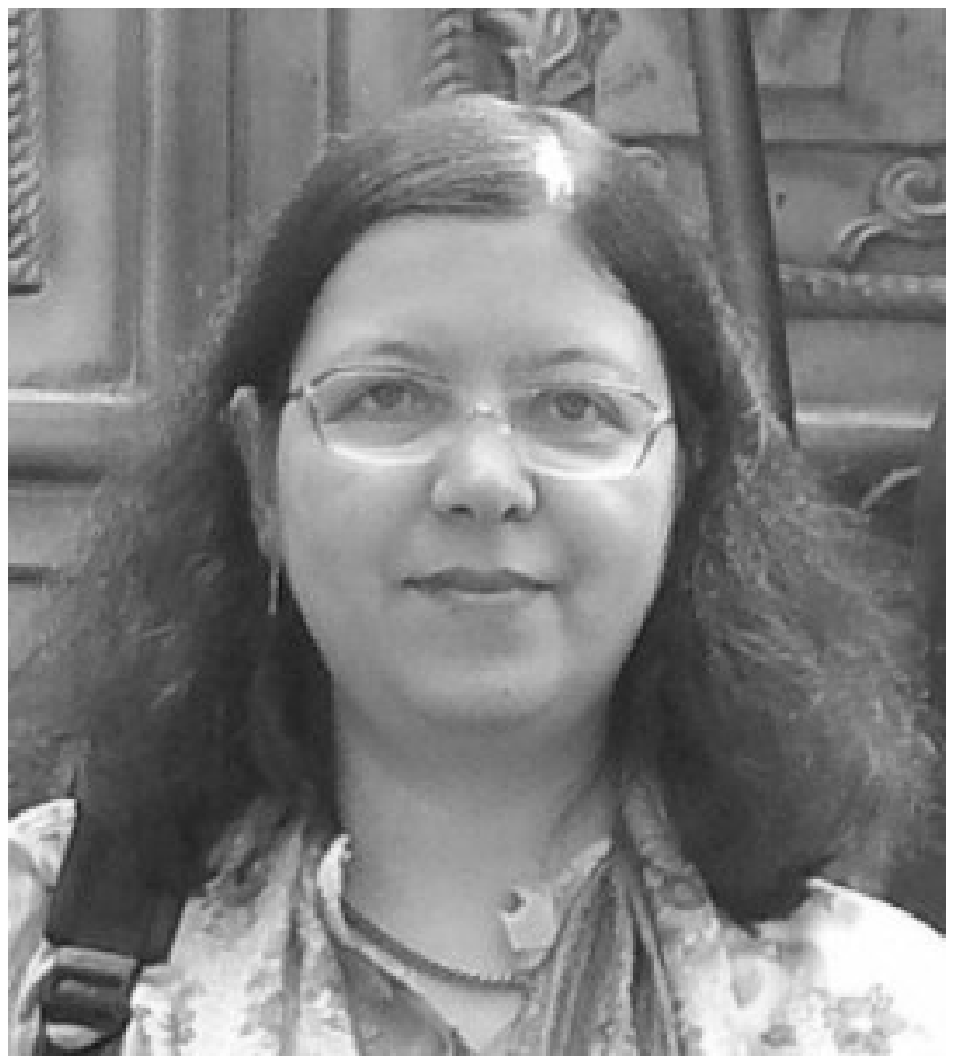}}]{Maryam Shojaei Baghini}
(M'00 - SM'09) received the M.S. and Ph.D. degrees
in electrical engineering from Sharif University of Technology, Tehran, in
1991 and 1999, respectively. She has worked for more than 2
years in industry on the design of analog ICs. In 2001, she joined
Department of Electrical Engineering, IIT-Bombay, as a Postdoctoral
Fellow, where she is currently a Professor. She is the author/coauthor of
143 international journal and conference papers, the
inventor/coinventor of 6 granted US patents, 1 granted Indian patent and
25 more filed patent applications.  Her current research interests include
high-speed links for on-chip and off-chip data communication,
high-performance low-energy analog/mixed-signal/RF IC design and test for
various applications including healthcare and sensor networks,
device-circuit co-design in emerging technologies and energy harvesting
circuits and systems. Dr. Shojaei serves in the Technical Program
Committee of several IEEE conferences, including IEEE International
Conference on VLSI Design, and Asia Symposium on Quality Electronic
Design. She was a TPC member of IEEE-ASSC from 2009 to 2014. Dr. Shojaei
is joint recipient of 11 awards
and a senior member of IEEE.
\end{IEEEbiography}
\vfill
\begin{IEEEbiography}[{\includegraphics[width=1in,height=1.25in,clip,keepaspectratio]{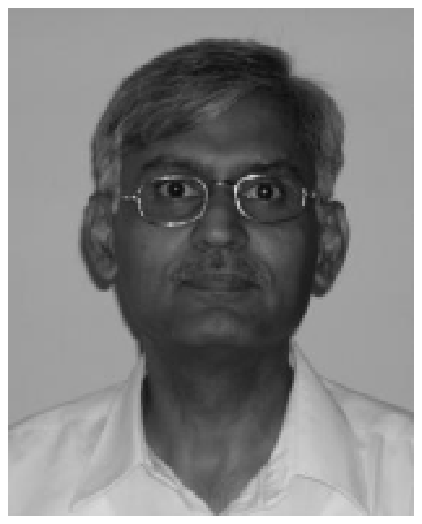}}]{Dinesh Sharma} 
obtained his Ph.D. from the Tata Institute of Fundamental
Research, Mumbai. He has worked at TIFR and IITB at Mumbai, at LETI at Grenoble in
France and at the Microelectronics Center of North Carolina in the U.S.A. on MOS
technology, devices and mixed mode circuit design. He has been at the EE deptt. of IIT
Bombay since 1991, where he is currently a Professor. His current interests include mixed
signal design, interconnect technology and the impact of technology on design styles.
\\
He is a senior member of IEEE, a fellow of IETE and has served on the editorial board of
"Pramana", published by the Indian Academy of Science.
\end{IEEEbiography}

\vfill
%\enlargethispage{-5in}
\end{document}